# Embedding Equity, Diversity, and Inclusion in the WST Collaboration

Laurane Fréour[a], Anna Puglisi*[b,c], Maria Cristina Fortuna[c], Fatemeh Zahra Majidi[d], Umberto Rescigno[e], Amelia Bayo[f], Francesca Primas[f], Sabine Thater[a], Laurence Tresse[g], Stephanie Escoffier[h], Letizia P. Cassarà[i], Roland Bacon[j], Sofia Randich[c], Vincenzo Mainieri[f]
[a]University of Vienna; [b]University of Southampton; [c]INAF - Osservatorio Astrofisico di Arcetri; [d]INAF - Osservatorio Astronomico di Capodimonte; [e]Instituto de Astronomía y Ciencias Planetarias, Universidad de Atacama; [f]European Southern Observatory; [g]Laboratoire d'Astrophysique de Marseille; [h]Aix Marseille University, CNRS/IN2P3, CPPM, Marseille, France; [i]INAF IASF Milano; [j]Centre de Recherche Astrophysique de Lyon

**Abstract**

The proposed Wide-field Spectroscopic Telescope (WST) is a next-generation telescope facility with a 12-meter primary mirror that may be operational in the 2040s. The project has a broad range of scientific goals and engages a large Community of more than 1000 members across multiple time zones, institutions, career stages, and professional roles. As for any large-scale scientific collaboration, building WST is therefore not only a technical and scientific challenge, but also a social and organisational one. This is particularly important in Astronomy and STEM disciplines more broadly, where marginalized groups remain underrepresented, particularly in leadership positions.

In this paper, we describe the creation and first year of activity of the WST Equity, Diversity, and Inclusion (EDI) working group. We discuss how the group has worked to embed EDI considerations into the early phases of the Collaboration, and we present our ongoing and future projects aimed at broadening participation, accountability, and community sustainability, while delivering groundbreaking science. We also reflect on the challenges of sustaining EDI work in a large and rapidly evolving collaboration, including recognition of EDI contribution, uneven participation, communication across heterogeneous groups, and the need to connect facility-level initiatives with broader organisational and territorial responsibilities. We present our activities as a situated example of how EDI can be approached as part of the governance, communication, and community infrastructure required to build sustainable research environments.

# 1. EQUITY, DIVERSITY, INCLUSION AND SUSTAINABILITY IN ASTROPHYSICS

The scale of modern astrophysics implies long-term international ecosystems sustained by collaborations that involve hundreds, sometimes thousands, of people across dozens of institutes worldwide. This makes the field deeply interdisciplinary, international, multicultural, as well as increasingly relational and interdependent. In this context, Equity, Diversity, and Inclusion (EDI) are essential core values for the successful development and implementation of new projects.

Equity is the acknowledgement of differences among individuals, with the goal of redistributing resources to achieve an equal outcome; Diversity refers to the representation of multiple identities; Inclusion means welcoming and valuing the thoughts, ideas, and perspectives of every individual. Because multiple identities and specificities exist in large multicultural and diverse teams, it is also important to consider that differences are not isolated, and each individual can identify with multiple marginalized categories. Therefore, including EDI values in scientific projects and ultimately fostering a culture of belonging that enables each individual to contribute their full potential requires an intersectional approach. Intersectionality is a framework for understanding how overlapping identities (e.g., race, gender, class, sexuality, and ability) interact to create unique systems of discrimination or privilege. In this context, EDI principles are also closely tied to sustainability, since a truly resilient future requires balancing the health of the planet with the well-being and welfare of all its people. While sustainability often focuses on environmental preservation, its goals cannot be achieved without the "Social" pillar of the Environmental, Social, and Governance (ESG) framework, and EDI practices can ensure that green initiatives are equitable rather than exclusionary.

*Corresponding author: Anna Puglisi (a.puglisi@soton.ac.uk)

While diversity has been recognised as a factor in improved company performance[1, 2], and companies strive to embed sustainable practices within their business (see, e.g., Ref. 3) responding to consumers' and investors' expectations, many obstacles remain on the path to EDI, even at the very early stages of entering the labor market. For example, people with disabilities and foreign-born women tended to be 24% and 18% less employed than people without disabilities and foreign-born men within the European Union in 2024, respectively[4]. Challenges extend way beyond employability and are also related to the every-day reality that individuals experience within their working environment, ultimately affecting how meaningfully they can engage and contribute. Complete statistics on discrimination and biases across academic careers are unfortunately less available. However, gender balance analysis in the European Union in 2022 already shows that only 33% of individuals in Grade A positions (i.e., Director’s level) are women[5]. Focusing on the UK and Ireland physical sciences sector, there is a general lack of awareness of LGBT+ issues in the workplace, and 28% of LGBT+ respondents had at some point considered leaving their workplace because of the climate or discrimination towards LGBT+ people[6]. The Astro2020 decadal survey highlights a lack of racial and ethnic diversity in US astronomy faculties, with African American and Hispanic people being up to 8 times less represented than in the US population[7]. The same survey report also notes how astronomers have not always succeeded in including local and native communities in the decision-making process for new telescopes and facilities, as demonstrated by the protests against large observatories at Mauna Kea, which ultimately led to delays and uncertainties in the construction of the Thirty-Meter Telescope project.

These studies provide valuable insight into representation, discrimination and biases within the labor market, and highlight the limitations in the current discourse around EDI and sustainability in Astrophysics. However, the ambitions of modern Astrophysics projects demand to reflect collectively on the environmental, social, and economic context of our science and on its implied responsibilities. Indeed, Astrophysics’ discoveries have consistently demonstrated the capability of inspiring wonders in the general public[8], and therefore attract new talents towards STEM disciplines[7]. Also, modern telescopes require increasingly ambitious technologies, thus driving significant progress and technological advances within society[9] and are often built in fragile climate ecosystems. Their development requires careful consideration of environmental impacts and the meaningful integration of local and indigenous communities into the decision-making process. Finally, as Astrophysics projects often unfold over decades, it is all the more vital to carefully evaluate their impact on future generations, also considering that the enormous scale of the related research communities risks perpetuating existing inequalities if these are not properly accounted for.

In this paper, we discuss our experience embedding EDI within the Wide-field Spectroscopic Telescope (WST)[10], to ensure that its potential for advancing knowledge and technological innovation is fulfilled while supporting the growth and inclusion of everyone involved. The goal is not to present WST as a finished model, but to document an early-stage attempt to embed EDI within the Collaboration governance. The paper is organised as follows. We describe the WST-EDI working group in Sect. 2, our ongoing projects in Sect. 3 and planned activities in Sect. 4. We briefly discuss challenges and lessons learnt in one year of activities in Sect. 5. Finally we present our conclusions in Sect. 6. Throughout the paper, “the WST Consortium” refers to the research institutes and universities included in the facility[1] concept study whereas “Science Team” refers to the broader network of professional astronomers that is developing WST's science drivers[2]. Finally, we use the term "WST Collaboration" to indicate collectively both the WST Consortium and Science Team.

[1] https://wstelescope.eu/horizon-consortium/

# 2. THE WST EDI WORKING GROUP

### 2.1 Group creation and composition

The WST is a concept study for a next-generation telescope, entirely dedicated to spectroscopic surveys. Combining a large field of view, a high-multiplex and a giant integral field spectrograph, it will collect hundreds of millions of spectra of galaxies and stars, revealing the Universe in unprecedented detail[10,11]. This ambition demands a parallel effort of engaging a broad and diverse community of astrophysicists in its operations. The EDI group emerged at the end of 2023 from a community-driven bottom‑up process, following a series of discussions with key leaders of the WST consortium. Doing so at the early stages of this Collaboration also provided a perfect opportunity to lay the groundwork for integrating EDI principles at the core of its decision-making activities and team culture.

The idea was warmly supported by management, and this has so far been fundamental for the successful establishment of the group, the delivery of key activities, and smooth operations grounded in a climate of mutual trust. The WST Project Office responded to this community-driven initiative by creating a dedicated working group within the successful European Union's Horizon Europe proposal to fund a conceptual study of the telescope. Including the EDI group as a stand-alone working group within the "General Facilities" work package was an important step for formal recognition of the group's activity and credit to its structural importance for the operations of the telescope. Moreover, this contributed to establishing EDI and governance accountability early in the Collaboration.

The EDI lead and Project Office opened a call for working-group members via a sign-up form shared with the general mailing list of the WST community, and the group officially began its activities in April 2025. We encouraged involvement from people with a wide range of backgrounds across all areas of the Consortium (Science, Hardware Design & Interface, Operations and General Facilities) and Science Team. No previous experience with EDI was required. At present, the group mostly consists of members from the Science Team and predominantly women based in European institutes, reflecting the challenges in engaging with specific sectors of the Consortium and broader Community (see Sect. 4 and 5). Members of the group span a broad range of career stages, with representation from senior and postdoctoral researchers as well as early‑career staff. This helps us include the voices of WST's most vulnerable members while at the same time engaging effectively with the Project Office and Steering Committee around key governance discussions. The group elected a co-lead via an anonymous vote at the very beginning of its activities. The EDI co-leads share the organization of the group's activities and contribute equally to the decision-making process to support the inclusion of multiple voices and perspectives.

### 2.2 Role of the group and Code of Conduct

The EDI working group was established with the goal of providing a consistent Point of Contact for the global WST community and facilitating communication, identifying emerging needs, and ensuring that concerns or suggestions are heard and addressed. Furthermore, it serves as a central advisory body that supports, informs, and collaborates closely with the Project Office on governance matters. The WST is a young and rapidly evolving collaboration: the EDI group is designed to evolve in response to community input and project requirements.

---

[2] https://wstelescope.eu/science-team/

To support the continued and sustained EDI activities through the whole life-cycle of the project, we set a number of shared core values that guide behaviour, collaboration, and decision-making across the WST Community. These values articulate a collective and personal responsibility to maintain a work environment grounded in trust, professionalism, mutual respect, integrity, honesty, transparency, and care. The Code of Conduct[12] was then developed as a practical tool to translate these values into shared expectations, guidance, and accountability mechanisms with support from the Project Manager and Project Office, and with input from the Steering Committee in the final stages of the document development. This document is viewed as a tool to empower every member of the wide and diverse WST Community to engage and contribute fully by setting shared expectations and by providing a practical guide for working across cultures, career stages, and communication styles. The Code of Conduct also contains a practical "how-to" guidance for expected behavior in every WST context. This includes face-to-face and online meetings, conferences and workshops, slack channels, workshops, or even social settings connected to our work, as well as examples of inclusive practices for meetings' organisation and coordination. The document explicitly discusses prohibited behavior and not-tolerated forms of harassment and discrimination to provide a reference for potentially affected individuals or bystanders, and facilitate recognition of accidents/complaints. In this context, the Code of Conduct also outlines a clear Accountability, Reporting and Complaints Resolution Procedure to support members of the community facing difficult situations. To this end, it also describes the role and responsibility of Ombudspersons (see also Sect 2.3 below).

The Code of Conduct, formally adopted on July 17th 2025, has since been actively disseminated and advertised via conferences, presentations, and emails. The document is available and clearly visible on the WST website and on the Consortium Slack platform. While upholding its principles is a collective responsibility, WST leads and meeting organisers play a special role in facilitating compliance with the Code of Conduct, and are therefore expected to be familiar with its content, surface concerns when appropriate, help route issues, and include regular reminders in recurring agendas. To support clear and accessible dissemination of our Core Principles and Conduct in various settings, a one-page summary as well as a one-slide version of this summary have been made available (see Figure 1). The latter in particular is intended for use during meetings and events to ensure shared understanding and consistent application of our values across the Collaboration.

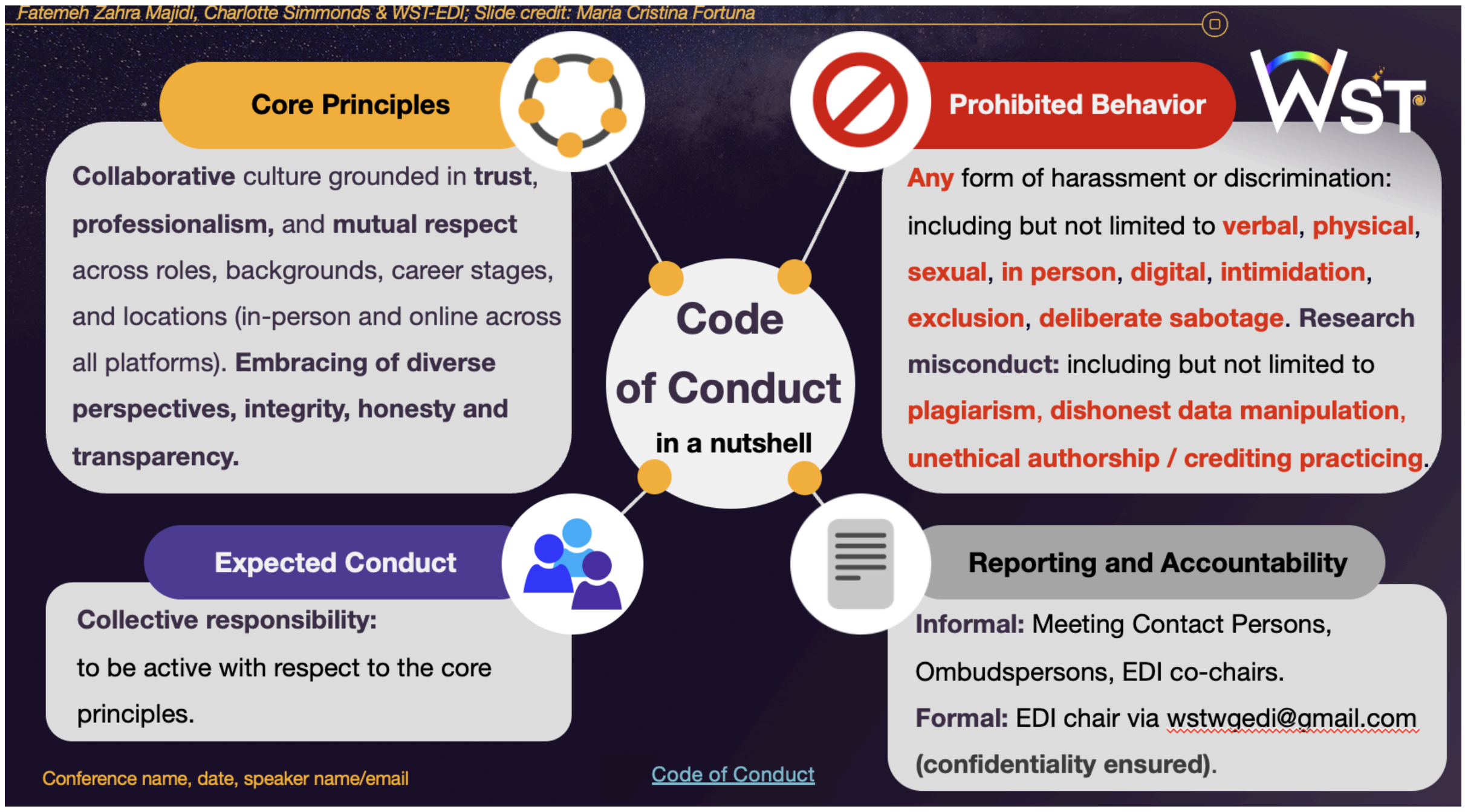

*Figure 1: One-page summary of the WST Code of Conduct. Credit: Fatemeh Zahra Majidi, Charlotte Simmonds, Maria Cristina Fortuna and the WST EDI and Communication working groups.*

### 2.3 Ombudspersons[3]

To ensure that individuals experiencing conflicts or challenges are supported in an independent and fair way, WST members elected three Ombudspersons with a proven record of commitment to EDI, experience in mediation and active listening. The nominations were reviewed and vetted by the EDI group, also following discussions with the Project Office. The Ombudspersons are expected to serve the WST Community for a renewable three-year term via an open call for nominations from the Community. Ombudspersons' contact information is easily accessible to the community via the EDI section of the WST website. They are expected to operate under the core principles of confidentiality, impartiality, independence from WST leadership, and informality when offering informal support for members experiencing interpersonal, group, or professional challenges within the context of WST activities, in line with the standards of the International Ombuds Association[4].

## 3. ONGOING PROJECTS

### 3.1 Inclusive Leadership Course

To support the full integration of our core values and Code of Conduct principles within all activities of the WST, it is crucial to build awareness and familiarity with EDI themes up to its leadership roles. This is particularly relevant because leaders have a significant influence on how people interact day-to-day[13] and must champion compliance with the Code of Conduct (see Sect. 2.3), while individuals with extensive EDI expertise are often underrepresented in leadership roles, a structural gap widely documented both in European institutional strategies[14] and international management literature[15,16]. Targeted management training can mitigate demographic biases and cultivate an inclusive climate[17].

The EDI group, in collaboration with the Project Office, is therefore designing an interactive session focused on tools for managing discussions, decision-making, and participation in large international teams, with the support of an external expert in EDI management and leadership. The learning objectives of this course have been co-designed with the course provider to ensure that the course content is tailored to the specific needs of the WST Collaboration. The workshop is aimed at WST members across all levels of leadership, from (sub-)working-group coordination to the Steering Committee. Participation will be capped at around 25 people to keep the format effective and maximise engagement. Depending on funding and the provider's availability, we will explore the possibility of organising a second session to allow more people to participate. The course is expected to be offered in the second half of 2026.

### 3.2 An EDI approach to communication

To diversify access and participation in the WST Collaboration, we are implementing an EDI approach to our communication strategy by actively collaborating with the Communication working group. Communication and EDI are closely connected areas of work: both deal with access, participation, visibility, and belonging, although with different scopes and focuses. In this sense, communication is not only how a collaboration presents itself internally and externally, but also how members find information, understand how the Collaboration is structured, and identify possible entry points for participation.

The WST communication is currently primarily in English and mostly oriented towards the scientific community, which comprises people from multiple backgrounds, including many non-native English speakers. In this context, while part of the language barrier is reduced by the shared scientific background, important differences remain due to career stage, specific field, role within the Collaboration, and familiarity with the WST structures. To embed EDI principles in communication, the WST Collaboration is currently focusing on three main routes that we describe below.

[3] We intentionally use the term "Ombudspersons" rather than the more familiar "Ombudsman" to favour gender-neutral language.

[4] https://www.ombudsassociation.org/

### 3.2.1 Resources

To support accessible communication throughout the Consortium, the Communication team adopted a strategy that facilitates implementation through built-in resources. These include slide templates designed to meet accessibility criteria[18], such as highly legible fonts, appropriate line spacing and paragraph structure, adequate background-foreground colour contrast, and a suitable palette for colourblind users. The Communication team also provide guidelines on how to make accessible presentations, including instructions on how to insert alt text in PowerPoint presentations, and a list of online tools, such as WebAIM contrast checker[5] and Colblis[6], to check whether materials satisfy accessibility requirements. Finally, the Communication team is producing a set of infographics and graphic material, each accompanied by its own alt text.

Despite these efforts, one of the challenges remains how to balance the need to produce accessible, yet highly technical material with the speaker's freedom and the practical constraints of scientific communication. Much of the graphical material is produced directly by members of the Collaboration and is often already published, leaving little room to improve its accessibility. We also acknowledge that fully accounting for accessibility considerations can require a significant investment of time. In many cases, speakers do not have this extra time, and the support that the Communication team can provide on an individual basis is resource-limited. One of our planned next steps is thus to provide proper training on how to create accessible plots and graphic material for consortium members, in order to facilitate their work and help recognise the value added by this effort.

The Communication and EDI working groups are also currently exploring practical solutions to provide the WST Collaboration with clearer guidance within non-standardised and heterogeneous contexts. A possible by-product of this effort could be a set of publicly available guidelines and resources.

### 3.2.2 The WST Chronicle

The WST Chronicle[7] is a three-monthly magazine distributed to the Consortium and wider WST Science Team, as well as published on the website. This magazine provides an in-depth view of selected topics related to the progress of the WST infrastructural operations (e.g., site selection, detectors), events (conferences and workshops), and Collaboration composition. In both the Chronicle and the website (see below), the Editor and Project Office have aimed to balance technological and scientific content, and to represent the work of different groups within the Collaboration. The Chronicle features two or three interviews per issue, with attention given to the alternation of topic, career stage, and gender of the people interviewed, although we note that a more systematic approach will require demographic data collection through a community survey (see Sect. 4.3). In this sense, the Chronicle is not only a communication tool, but also a way to publicly recognise individuals' contributions towards building the WST and make the Collaboration more visible to itself. Furthermore, with the Chronicle, the WST Consortium actively seeks to provide role models across career stages, identities, and cultural backgrounds, increasing the sense of belonging within the Collaboration itself and supporting its members to participate meaningfully rather than being solely represented.

### 3.2.3 The WST website

The website[8] provides a stable point of access to information about the project. It presents each topic in a comprehensive way, provides a clear map of the Consortium and Science Team organisation, and helps make the structure of the work more transparent and easier to navigate. Each page of the website is also linked to relevant articles from the Chronicle that provide the reader more in-depth materials on a given subject. The website is frequently updated and, in the near future, will also host a more detailed representation of the Collaboration behind the project and specific Collaboration highlights.

---

[5] https://webaim.org/resources/contrastchecker/
[6] https://www.color-blindness.com/coblis-color-blindness-simulator/
[7] https://wstelescope.eu/the-wst-chronicle/
[8] https://wstelescope.eu/

A key aspect of the website work is an active effort to make communication accessible, ensuring that different sensory and cognitive needs are accounted for. This supports independent access to resources and opportunities, enabling individuals to engage meaningfully with its content with minimal dependence on others. Interestingly, the benefits of accessibility extend beyond specific disability groups and can support inclusion at all levels. For example, it has been demonstrated that approaches originally developed for users who are visually or hearing impaired, or who are neurodivergent, often improve usability for many others, including older users, people using low-quality devices, non-native speakers, and users experiencing temporary or situational impairments[19]. This demonstrates how implementing EDI guidelines in communication is a strategy that can enhance and broaden participation.

Regarding website accessibility, although the Web Content Accessibility Guidelines (WCAG) principles were closely followed and the current digital world offers a number of assistive technologies, we faced a lack of standardisation. We invested significant time in testing the website beyond the simple built-in tools available in most internet browsers. Specifically, we ran automatic inspector software[9], which performs best for testing the overall structure, the usage of Accessible Rich Internet Applications (ARIA), and missing links or link labels, but is not designed for content-specific issues. We also used WAVE[10] (Web Accessibility Evaluation Tool), which provides more detailed feedback but requires an in-depth analysis of the output, resulting in a more time-consuming inspection. Finally, we performed tests with real users. In particular, we tested the performance of the website with a voice-over tool, and repeated the test with a blind first-time website user. This latter test was crucial not only because it identified issues that had been missed by automatic tools and basic audits, but also because it provided direct insight into how the website is actually navigated through assistive technologies. This revealed aspects of accessibility that are difficult to evaluate through checklists, automatic tools, or separate design and software checks, especially when the website is experienced as an integrated system.

### 3.3 Dissemination Plan

EDI structures and activities rely on strong engagement, trust, and inputs across all levels of a collaboration. Furthermore, structurally embedding EDI into the work culture of a large collaboration such as the WST requires the group to be visible, repeatedly named, and integrated into ordinary communication channels. Easy access to the group is also essential for providing support effectively to people facing difficult situations. Within the WST, we are therefore treating dissemination as part of the infrastructure of participation and we are developing and maintaining a set of communication strategies to make EDI work findable, accountable, and connected to the Collaboration. This has so-far included:

- Making EDI channels visible and accessible, through a dedicated EDI section on the WST website (see also Sect. 3.2), clear links to relevant documents, contact information, resources and complaints resolution procedures, as well as a dedicated article introducing the EDI group in the WST Chronicle issue 3. Where possible, we included pictures of group members/leads to support a sense of closeness, helping EDI structures appear less detached and more approachable;
- Embedding EDI into ordinary collaboration practices and normalizing our values of shared responsibility by including the one-slide version of the Code of Conduct into meeting and presentation templates;

[9] https://www.w3.org/WAI/test-evaluate/tools/list/ for a list of available resources
[10] https://wave.webaim.org/

- Maintaining regular communication with project structures, including regular engagement with the Project Office with openly-available minutes and agenda items, and sharing regular group activities and recording, minutes, and updates of EDI meetings via the WST Consortium Slack;
- Bridging EDI into scientific and technical spaces, through regular and consistent presence of the EDI group at the WST or WST-related conferences. Where possible, given the EDI group's capacity, team members give presentations about our activities. When direct participation from the EDI group is not possible, we actively reach out to meeting organisers and request a brief advertisement of the group. This helps attract technical professionals and backgrounds that are currently under-represented within the EDI group itself.

This visibility work is supported by the active cross-presence of members from the Communication and Sustainability working groups. While these areas remain distinct, they are closely interconnected: inclusive communication helps make EDI structures visible and accessible, and EDI perspectives help communication practices actively counter exclusion rather than reproduce it. Sustainability also intersects with EDI through questions of climate justice and unequal vulnerability. This cross-membership helps avoid treating EDI as an isolated add-on, and facilitates synergies across the Collaboration.

Beyond the WST, we also aim at connecting with the wider community to reach and engage individuals from traditionally marginalised groups, and to promote a collaborative scientific and operational model based on sharing knowledge, good practices, and lessons learnt. To this end, we have delivered invited and contributed talks at conferences[20], institute seminars, and astronomers networks, also within contexts that are not normally dedicated to EDI spaces to raise awareness and EDI capacity within the wider community. We have also submitted a White Paper[21] to the ESO Call supporting the Expanding Horizons initiative, and disseminated it through different channels, including outreach articles towards the general public[22]. The ESO White Paper advocates for embedding equitable governance structures as integral design parameters for the next flagship observatory, and provides an example of how internal EDI work can contribute to broader conversations around governance and future infrastructure. It was deliberately submitted to a call otherwise focused on scientific challenges to underscore the vital connection between equity and scientific excellence as the structural role of EDI and community work. This “EDI outreach” strategy supports our broader goals of making EDI work visible, promoting accountability, and cultivating a wider community connecting EDI groups across scientific collaborations and beyond (see also Sect. 5.5).

# 4. FUTURE PROJECTS

## 4.1 EDI Has a Gender: Mapping Participation and Labour in the WST EDI Group

The composition of our EDI working group provides an important starting point for reflection. While the group has benefited from the commitment of highly engaged members, participation has not been evenly distributed across the Collaboration. Some profiles, roles, and demographics remain less represented within the group. This limits the range of perspectives shaping the work and increases the risk that EDI work remains concentrated among those already most invested in these issues. This is not a WST-specific concern. A substantial body of literature provides a strong foundation for examining how EDI-related work is distributed within scientific collaborations, and identifies recurring tensions in its implementations. Men’s engagement in EDI work, especially in academic and organisational contexts, is often more limited, more conditional, and less stable than would be required for structural change[23]. Men may be more likely to support gender equality when they recognise women’s disadvantage[24], but this support can be constrained by fear of backlash, defensiveness, perceived status threat, or the perception that EDI work is not their responsibility[25]. Organisational studies further suggest that men in leadership or managerial roles are often positioned as central stakeholders in institutional change, while existing norms around leadership, status, and professional identity can make it difficult to recognise or redistribute community-sustaining work[26]. The literature also highlights the importance of invisible emotional and organisational labour, which often remains unrecognised despite being essential to the functioning of institutions[27]. At the same time, engagement dynamics may also be shaped by how accessible EDI spaces are to people who are not already fluent, trusted as legitimate contributors, or legitimate within them.

Within the WST EDI group, we would like to investigate the conditions that enable participation in EDI-related work through the project “EDI Has a Gender: Mapping Participation and Labour in the WST EDI Group”. The purpose of this project would be to understand the conditions that enable or discourage participation in EDI work, rather than assigning individual responsibility. In this sense, the project would examine how factors such as career stage, role within the Collaboration, perceived relevance of EDI, time availability, confidence, recognition, fear of saying the wrong thing, or perceptions of leadership responsibility shape engagement. This is particularly important because the work required to address EDI often includes visible tasks such as drafting documents, organising meetings or information activities, and preparing funding proposals, but also invisible tasks such as explaining EDI concepts, raising awareness, listening to concerns, mediating tensions, supporting excluded individuals, making overlooked problems visible, and repeatedly clarifying why these issues matter. When this labour is unevenly distributed, it can become a hidden workload for the same people [e.g., Ref. 28], while limiting the collective benefits of EDI practices.

We are currently planning to explore these dynamics through a mixed-method approach. First, an anonymous survey would gather broad information on participants’ views, willingness to engage, perceived obstacles, familiarity with EDI-related topics, and perceived responsibility for supporting EDI work. In parallel, semi-structured interviews would explore more deeply the attitudes, hesitations, motivations, and cultural or institutional barriers that shape participation. The interviews’ main value would be to move beyond general trends and better understand how individuals interpret their role, what encourages or discourages their engagement, and how their specific institutional or personal situation affects their participation. When designing this project, collaboration with professionals in the social sciences sector would help mitigate the impact of defensiveness, excessive formality, or social desirability in responses, and would support data interpretation and question design (see also Sect. 4.3 below).

This project would help us identify who participates in EDI work and under what conditions that participation is possible and meaningful. Over time, survey insights and data could help us design strategies to share EDI responsibilities more equitably, and broaden engagement. The findings could also bring information into whether EDI commitments are being implemented or are at risk of becoming largely symbolic, as highlighted in recent discussions about diversity and gender washing[29,30]. Ultimately, this would reduce the risk that community-sustaining work remains invisible or concentrated on a selected few, thus countering the paradox where inclusion may rely on an unequal distribution of effort.

### 4.2 Seminars

The EDI Working Group also aims to establish a regular seminar series to provide a foundational reference point for EDI within the Collaboration. These seminars will focus on various topics such as mental health, gender balance, and the practical implementation of equity and equality across diverse institutional landscapes. The goal is to act as a space for members to connect, share ideas, learn about a diversity of topics, and strengthen the Collaboration around EDI principles.

We will build these seminars on a successful pilot model implemented by the WST Time-Domain working group: short, informal talks by a mix of researchers, followed by an open discussion. This model has already demonstrated a unique capacity to foster synergies and catalyze collaborations both within and outside the consortium. By expanding this framework to EDI activities, we intend to actively engage the Collaboration and encourage interdisciplinary cooperation. To ensure the content remains responsive to the Collaboration’s evolving needs, open communication via e.g. regular polls will be essential to identify and prioritize the topics most relevant to our members.

### 4.3 Broadening horizons through listening: data, networks, and local communities

We have identified three additional and complementary future directions, still to be developed, that can strengthen the internal EDI structure of the WST Collaboration while opening dialogue with the wider astronomy and astrophysics one. These include listening and communication strategies: a community survey, which provides internal and structured listening; networking with other EDI groups, to support cross-collaborations communication; and dialogue with local and Indigenous communities, fostering listening towards those whose territories, expertise, and relationships will shape the physical reality of the infrastructure. Together, these directions can help ensure that the EDI working group does not remain only an inward-looking structure, but becomes part of a broader ecosystem of accountability, learning, and exchange.

A well-planned community survey can be a useful tool to connect with the Collaboration, assess individual well-being, and give members a valuable opportunity to express discomfort and raise issues. Large participation would allow us to gauge the general climate and culture across the Collaboration, as well as identify emerging systemic issues[31]. Finally, it would enable the collection of demographic and professional data, and hence help us understand how the diversity of the Collaboration is represented vertically across WST structures up to leadership positions. If resources allow, the community survey should be implemented with experts in qualitative and quantitative methods for the social sciences. Involving social-science professionals would support the formulation of questions in the most appropriate way. Indeed, an ill-defined question might produce meaningless data, flatten complex human experiences into numbers, or, in the worst-case scenario, lead to resentment and mistrust. Collaboration with social-science professionals would also support a methodologically sound and context-sensitive interpretation of the data, helping to avoid overly simplistic readings of quantitative indicators, false reassurance from aggregate numbers, or conclusions that overlook the specific work cultures and power dynamics of the Collaboration. This is especially important in scientific contexts where quantitative indicators are often assumed to be straightforward to interpret. The community survey would not be a one-off exercise, but would need to be repeated regularly every few years. This time-monitoring aspect would support assessing the impact of EDI work and ensuring that Collaboration diversity is reflected across decision-making, leadership, authorship, visibility, and governance levels, and thus better integrated within the fabric of the collaboration itself.

The issues that we tackle within the EDI working group are, for the most part, not a peculiarity of the WST context, but rather contextual manifestations of larger systemic issues that many other EDI groups are also discussing every day. Therefore, we would like to work towards building a network across EDI groups of scientific collaborations and beyond. This would allow us to share best practices, documents, strategies, and their limitations, while avoiding reinventing tools or repeating mistakes in isolation. Creating a wider EDI network would also allow us to “reclaim scientific and technical spaces” by bringing EDI themes into conferences, special sessions, publications, and scientific networks. This can help shift the tone of the conversation around scientific production, by highlighting the societal, infrastructural, and political conditions that make science results possible. In this sense, increasing the critical mass and visibility of EDI discourse within scientific spaces is itself an important goal. Not everyone will access or read EDI policies, but many people may encounter these questions through a talk, a session, a paper, or a discussion in a scientific context. Over time, this can progressively shift the baseline of what is understood as the minimal enabling conditions for better and more sustainable science. Finally, EDI work can often lead to a profound sense of isolation, as a result of confidentiality, limited time availability, the specificity of the issues addressed, and the slow response of scientific infrastructures (as of many societal infrastructures) to systemic change. EDI groups or individuals working in isolation risk loneliness and helplessness when facing the slowness of change. Creating networks, and leaning into them, can therefore be a form of emotional and political sustainability that actively counters frustration. Frustration can be less isolating when it is understood not as an individual failure, but as part of a collective experience of working against structural inertia. At the same time, such a network should not become an additional informal structure relying on the same voluntary and often under-recognised labour it aims to support, but rather a sustainable space for exchange, memory, solidarity, and shared responsibility.

Finally, we would like to support and advocate for meaningful engagement with local and Indigenous communities around the future WST site, recognising that such engagement must be led at the appropriate organisational level and should not be viewed as an isolated facility-level activity. Existing initiatives in the Chilean context provide important precedents for engagement, including archaeological documentation[32], preservation of Atacameño/Likan Antai cosmic worldviews[33], educational exchange[34], and support for Kunza language and cultural memory[35]. While the WST EDI group does not have the expertise or mandate to design this engagement independently, we believe it is important to ensure that this dimension is explicitly considered before major decisions and implementation pathways become fully stabilised. Otherwise, the risk is that activities connected with the territory are introduced only as communication, outreach, or compensation. Collaboration with local Chilean astronomers, community-informed partners, and, where appropriate, local and Indigenous communities themselves would be essential to work towards science that is co-created in connection with the territory. This would also require recognising that territorial relations also include environmental responsibility, labour conditions, and the material experiences of Chilean workers and local communities who make astronomical infrastructures possible. Early conversations can shape trust, legitimacy, environmental responsibility, and the social licence for scientific infrastructures to operate over decades. They also help move towards collaborative rather than extractive models of science, including the voices of people who are impacted by the existence of large infrastructures even when they are not formally part of the Collaboration.

# 5. Lessons and Challenges after one year of activities

In this section, we share challenges and lessons learnt after one year of activities. Rather than presenting these as final conclusions, we offer them as a starting point for further discussion and as a useful resource for other groups engaged in similar work.

To start, this year showed us that there are people eager to commit their time and effort for supporting the scientific community across various career stages, and that this does not necessarily require previous EDI experience. Instead, people can contribute from different backgrounds provided that there is willingness to listen, learn, and engage responsibly. These diverse perspectives can enrich the conversation and help highlight needs that might otherwise remain unnoticed. Also, this year demonstrated how important it is to establish a collaborative, yet critically engaged relationship with the main leadership structures of the Consortium: one that aims to support the project, while retaining enough independence to raise questions, identify gaps, and propose changes. At the same time, we noticed the following main challenges:

- Formally recognising EDI contributions to the Collaboration is not trivial. While some existing collaborations have already implemented some reward mechanisms for structural EDI work, it was not obvious to identify an appropriate compensation mechanism that could translate into tangible professional recognition, especially at such early stages, when various details of the facility operations are still under discussion. On the other hand, having a reward system is important to protect the career progression of EDI team members, who are often early-career researchers on temporary contracts, by preventing their time investment from becoming an invisible professional cost. At present, we have negotiated the inclusion in the WST publication policy document of a recommendation that EDI group members should be invited as co-authors on publications, provided that they meaningfully engage with them, for example by providing comments. This is not ideal because, for instance, it does not easily apply to working groups where publications are not a standard output or form of recognition, but it nonetheless represents a starting point that can and will evolve as the collaboration progresses.

- Our initial setup did not account for dedicated financial support for the EDI working group. However, our experience has highlighted that structured funding is vital for training as well as to onboard an EDI Officer or a professional with a background in social sciences. This dedicated expertise would allow for example to conduct rigorous surveys and successfully implement robust EDI strategies and governance structures.

- The WST Community includes more than 1,000 members across institutes, continents, and time zones, including engineers, technicians and software development teams, and working groups with diverse science goals. As discussed in Sect. 3.3, making the EDI group visible and building trust and a sense of belonging within such a complex collaboration is not obvious. This complexity is a challenge, and it is also a resource that we can learn to navigate with time. In this context, effective communication with the whole Collaboration remains an open question. A dedicated Slack channel exists, but this is limited to members of the Consortium. A mailing list including all members of the Science Team also exists, but it was created for science communication. We therefore need to find a balance between reaching as many people as possible and respecting the original scope of the list.

- Engaging beyond our immediate professional context or the identities currently represented within the EDI working group is not simply a matter of extending broader invitations. Even proactive engagement can reproduce blind spots. The language, formats, and assumptions used by an EDI group can still reflect the perspectives of its most active members, unintentionally creating distance from the very groups it aims to involve. In our case, this seems to be particularly relevant for engagement with technical and operational teams, which are underrepresented within the EDI group, but it also applies to broader demographics within the scientific community, including groups that might be overrepresented in leadership but underrepresented in EDI work (see also Sect. 4). After one year, we see that the EDI working group does not only need "better communication strategies" towards different groups, but also a critical assessment of its own assumptions and conditions for engagement. A large Collaboration such as the WST is not a homogeneous community, but rather a combination of professional cultures, constraints, priorities, levels of familiarity or trust with EDI language, and vulnerabilities. The challenge seems to be not only how to translate across professional contexts but also how to remain open about our blind spots and, where possible, co-develop structures that can be flexible enough to welcome multiple vocabularies.

- Sustained, long-term engagement with the activities of an EDI group is essential for building trust and acting effectively. Yet, time investment can be limited and highly variable over time, due to, e.g., career duties or personal responsibilities. This applies across working groups, but its impact within the EDI context may be stronger. Sustaining healthy scientific communities requires essential labour such as mentoring, conflict resolution, codes of conduct, and community care, which often happens on top of scientific duties. This work often falls disproportionately on early-career researchers and marginalised people. Furthermore, it can be emotionally demanding and is often undervalued in hiring, promotion, and evaluation processes, contributing to the "EDI penalty", whereby people who sustain the Collaboration become less visible as scientists. How to design EDI work so that its continuity does not depend on individual overcommitment, therefore remains an open challenge. Possible strategies for addressing it may include: clearer task distribution; better representation of diversity within the EDI group; rotating responsibilities; documentation and handover practices; formal recognition mechanisms; and clearer boundaries around the EDI group's remit.

# 6. CONCLUSIONS

In this article, we reported on the creation of the EDI working group in the WST Collaboration, which provides a situated example of how EDI can be approached as part of the governance, communication and community infrastructure within a large scientific collaboration. We discussed our ongoing and future projects, as well as the challenges we are facing in embedding EDI in a Collaboration of more than 1,000 members with diverse backgrounds, cultures, and experiences. While complex and challenging, however, incorporating EDI principles is essential for sustainable progress. It requires flexibility and constant evolution to account for rapidly evolving collaborations, but it is also important to have a long-term vision to ensure that short-term projects and goals are meaningfully incorporated into the broader picture.

## ACKNOWLEDGEMENTS

This project has received funding from the European Union Horizon Europe Research and Innovation Action under grant agreement no. 101183153 -WST. Views and opinions expressed are however, those of the author(s) only and do not necessarily reflect those of the European Union or the European Research Executive Agency (REA). Neither the European Union nor the REA can be held responsible for them. FZM acknowledges support from the ASI-INAF Agreement no. 2021-12-HH.0 and Addendum 2021-12-HH.1-2024 "Missione Solar-C EUVST-Supporto scientifico di Fase B/C/D".